%% file: main.tex
\documentclass{PoS}
\usepackage{multicol}
\usepackage{amsmath}
\usepackage{bbm}

\title{The gradient flow coupling from numerical stochastic perturbation theory}

\ShortTitle{The gradient flow coupling from NSPT}

\author{\speaker{Mattia Dalla Brida}\,\thanks{Current address:
       \emph{CERN, Theoretical Physics Department, 1211 Geneva 23, Switzerland}.
       The author is grateful to the Theoretical Physics Department at CERN
       for the hospitality and support.}\\
       NIC, DESY, Platanenallee 6, 15738 Zeuthen, Germany\\
       Dipartimento di Fisica, Universit\`a di Milano-Bicocca 
       \& INFN, sezione di Milano-Bicocca, Piazza della Scienza 3,
       I-20126 Milano, Italy\\ 
       E-mail: \email{mattia.dalla.brida@desy.de}}

\author{Martin L\"uscher\\
        CERN, Theoretical Physics Department, 1211 Geneva 23, Switzerland\\
        Albert Einstein Center for Fundamental Physics, 
        Institute for Theoretical Physics, \\
        University of Bern, Sidlerstrasse 5, 3012 Bern, Switzerland\\
        E-mail: \email{luscher@mail.cern.ch}}

\abstract{Perturbative calculations of gradient flow observables are technically 
	  challenging. Current results are limited to a few quantities and, in
	  general, to low perturbative orders. Numerical stochastic perturbation
	  theory is a potentially powerful tool that may be applied in this context.
	  Precise results using these techniques, however, require control over both
	  statistical and systematic uncertainties. In this contribution, we discuss
	  some recent algorithmic developments that lead to a substantial reduction
	  of the cost of the computations. The matching of the ${\overline{\rm MS}}$
	  coupling with the gradient flow coupling in a finite box with Schr\"odinger
	  functional boundary conditions is considered for illustration.\\
	  
	  {\vspace*{3mm}\hfill CERN--TH--2016--239}}

\FullConference{34th annual International Symposium on Lattice Field Theory\\
		24-30 July 2016 \\ University of Southampton, UK}

\input macros

\begin{document}

\input sect1.tex

\input sect2.tex

\input sect3.tex

\input sect4.tex

\input sect5.tex
\input acknw.tex
\input bibl.tex

\end{document}

%% file: macros
\def\proof{\noindent{\sl Proof:}\kern0.6em}

\def\frac#1#2{\hbox{$#1\over#2$}}
\def\dual{\mathstrut^*\kern-0.1em}

\def\lvec#1{\setbox0=\hbox{$#1$}
    \setbox1=\hbox{$\scriptstyle\leftarrow$}
    #1\kern-\wd0\smash{
    \raise\ht0\hbox{$\raise1pt\hbox{$\scriptstyle\leftarrow$}$}}
    \kern-\wd1\kern\wd0}
\def\rvec#1{\setbox0=\hbox{$#1$}
    \setbox1=\hbox{$\scriptstyle\rightarrow$}
    #1\kern-\wd0\smash{
    \raise\ht0\hbox{$\raise1pt\hbox{$\scriptstyle\rightarrow$}$}}
    \kern-\wd1\kern\wd0}
\def\slash#1{\setbox0=\hbox{$#1$}\setbox1=\hbox{$\kern1pt/$}
    #1\kern-\wd0\kern1pt/\kern-\wd1\kern\wd0}

\def\nabstar#1{{\nabla\kern0.5pt\smash{\raise 4.5pt\hbox{$\ast$}}
               \kern-5.5pt_{#1}}}

\def\drvstar#1{{\partial\kern0.5pt\smash{\raise 4.5pt\hbox{$\ast$}}
               \kern-6.0pt_{#1}}}

\def\ldrvstar#1{{\lvec{\,\partial}\kern-0.5pt\smash{\raise 4.5pt\hbox{$\ast$}}
               \kern-5.0pt_{#1}}}

\def\MSbar{\overline{\rm MS\kern-0.5pt}\kern0.5pt}

\def\zetabar{\bar{\zeta}}
\def\zetaprime{\zeta\kern1pt'}
\def\zetabarprime{\zetabar\kern1pt'}

\def\diracstar#1#2{
    \setbox0=\hbox{$\gamma$}\setbox1=\hbox{$\gamma_{#1}$}
    \gamma_{#1}\kern-\wd1\kern\wd0
    \smash{\raise4.5pt\hbox{$\scriptstyle#2$}}}

\def\Obs{{\mathcal O}}

\def\Ds{D_{\rm s}}
\def\DsdagDs{\Ds{\Ds}^{\kern-1pt\dagger}}

\def\avg#1{{\kern1.0pt\overline{\kern-1.0pt#1\kern-1.0pt}\kern1.0pt}}


%% file: sect1.tex
\section{Introduction}

The Yang-Mills gradient flow (GF) is a powerful tool to solve
renormalization problems in lattice QCD~\cite{Luscher:2010iy,
Luscher:2011bx,Luscher:2013vga}. Step-scaling studies, for example,
may be based on observables defined at positive flow time, and 
these are then related to standard renormalization conventions at
high energies using perturbation theory. The so-called small flow
time expansion of local fields is another instance, where perturbation
theory plays an important role. These perturbative computations 
however tend to be technically challenging and the calculations have 
so far been limited to low orders in the coupling and a restricted 
set of quantities (see e.g.~refs.~\cite{Luscher:2010iy,Fodor:2012td,
Fritzsch:2013je,Suzuki:2013gza,Bribian:2016}). Only recently a two-loop
computation was carried out~\cite{Harlander:2016vzb}. 

In this context, numerical stochastic perturbation theory 
(NSPT)~\cite{DiRenzo:1994sy,DiRenzo:2004ge} is a potentially useful 
tool. NSPT provides in principle a very general framework for high order
perturbative lattice computations of GF quantities~\cite{Brida:2013mva}. 
These techniques, however, come with some limitations, and it is not 
obvious that precise \emph{continuum} results are attainable in practice. 
In this contribution we intend to show that recent algorithmic developments
in this field, may indeed allow us for such precise determinations, at 
least up to two-loop order. 

In the next section we describe a new form of NSPT which is based on the 
stochastic molecular dynamics (SMD) equations. In Section \ref{sec:GFcoupling}, 
we introduce the specific observable we considered for this study, namely 
the GF coupling proposed in~\cite{Fritzsch:2013je}. In Section \ref{sec:Results},
we discuss our results for the one- and two-loop matching of this coupling and
the ${\overline{\rm MS}}$ coupling. Particular attention is given to the 
difficulties encountered in the continuum extrapolations.
We finally conclude in Section \ref{sec:Conclusions} with some remarks. 
Note that in this contribution we focus on the pure SU(3) Yang-Mills 
theory. The case of QCD will be briefly commented later on.

%% file: sect2.tex
\section{SMD based NSPT}
\label{sec:NSPT}

In its original form NSPT is based on the Langevin equations,
and amounts to solving the Parisi-Wu equations of stochastic
perturbation theory numerically~\cite{DiRenzo:2004ge}. From a 
numerical point of view, other stochastic differential equations 
might however lead to more efficient implementations of 
NSPT~\cite{Brida:2015d}. Here we consider the SMD 
equations~\cite{Horowitz:1985kd,Horowitz:1986dt}, which in the 
case of the pure SU(3) (lattice) gauge theory read (see e.g. 
ref.~\cite{Luscher:2011kk}),
\begin{equation}
  \label{eq:SMD}
  \begin{aligned}
  \partial_{s} U_{s}(x,\mu) &= g_0\pi_{s}(x,\mu) U_{s}(x,\mu),\\
  \partial_{s} \pi_{s}(x,\mu) &= -g_0\nabla_{x,\mu} S_G(U_{s})
  -2\mu_0\pi_{s}(x,\mu) + \eta_{s}(x,\mu),\\
  \langle\eta^a_{s}(x,\mu)&\eta^b_{r}(y,\nu)\rangle_\eta = 
  4\mu_0\delta^{ab}\delta_{\mu\nu}\delta(s-r)a^{-4}\delta_{xy}.
  \end{aligned}
\end{equation}
In the above equations, $\nabla_{x,\mu}S_G$ is the derivative 
of the gauge action $S_G$ with respect to $U_s(x,\mu)$, $g_0$ is 
the bare coupling, $a$ is the lattice spacing, and $\mu_0>0$ is 
a free parameter. As usual, $U_s(x,\mu)$ denotes the gauge field
and $\pi_s(x,\mu)$ the associated momentum field, while $\eta_s(x,\mu)$
is a random field with values in the Lie algebra of SU(3) and 
normal distribution. All fields depend parametrically on the 
\emph{stochastic (or simulation) time} $s$.

As is well known, the numerical solution of eqs.~(\ref{eq:SMD}) 
starts with the discretization of the stochastic time in units of 
a step-size $\delta\tau$, i.e. $s\to s=na\delta\tau$, $n\in\mathbb{Z}$. 
The discrete equations are then solved by alternating single steps of
molecular dynamics (MD) evolution, corresponding to eqs.~(\ref{eq:SMD})
with $\mu_0=0$, with a partial refreshment of the momenta. In a 
NSPT implementation, the main difference is that all operations 
involved are performed in a \emph{order by order} fashion. This means
that the fields are considered to have an expansion of the form,
\begin{equation}
\label{eq:Fields}
U_s(x,\mu)=\mathbbm{1}+\sum_{k=1}^{M} g_0^k\,U_{s,k}(x,\mu),\quad
\pi_s(x,\mu)=\sum_{k=0}^{M-1} g_0^k\,\pi_{s,k}(x,\mu),\quad
\eta_s(x,\mu)=\eta_{s,0}(x,\mu),
\end{equation}
and the theory is solved to a given order $M$ in the 
coupling.\footnote{In fact a stable numerical integration of eqs.~(\ref{eq:SMD})
in NSPT requires the inclusion of a gauge damping term~\cite{DiRenzo:2004ge,
Hesse:2013lat}. The discussion is however rather technical and is 
omitted here.} Once a Monte Carlo history of say $N$ such field 
configurations is generated, the perturbative expansion 
of a generic expectation value, $\langle\Obs[U]\rangle|_{g_0^M}=
\sum_{k=0}^{M}g_0^k\,c_k$, can readily be estimated. Explicitly, one 
simply computes the expansions $\Obs[U_s]=\sum_{k=0}^{M}g_0^k\Obs_{k}[U_{s,0},
\ldots,U_{s,k}]$ and averages the coefficients $\mathcal{O}_k[U_{s,0},
\ldots,U_{s,k}]$ over the simulation time, thus obtaining:
$\overline{\mathcal{O}}_k=(1/N)\sum_{n=0}^{N}
\mathcal{O}_k[U_{na\delta\tau,0},\ldots,U_{na\delta\tau,k}]$.
The coefficients $\overline{\mathcal{O}}_k$ are biased 
estimators of the perturbative coefficients $c_k$, in the sense that only
$\lim_{\delta\tau\to0}\,\lim_{N\to\infty}\,\overline{\mathcal{O}}_k=c_k$.

From the above discussion it appears clear that NSPT provides in 
principle a very general and compelling set-up for tackling 
challenging perturbative computations. As anticipated, however, 
these techniques suffer from some limitations. First of all, 
for a finite Monte Carlo sampling the perturbative 
coefficients $\overline{\mathcal{O}}_k$ come with a finite statistical
error $\sigma(\overline{\mathcal{O}}_k)$. Moreover, the algorithm suffers 
from critical slowing down, which makes the continuum limit difficult 
to approach at fixed statistical precision (s. below). Secondly, the 
algorithm is not exact: the coefficients $\overline{\mathcal{O}}_k$ 
have systematic O($\delta\tau^p$) errors, where $p$ is the order of
the integration scheme employed for the MD steps. These need to be 
extrapolated away, or rather $\delta\tau$ has to be chosen small 
enough for these effects to be negligible with respect to the target
accuracy.

On the other hand, the SMD algorithm has a free parameter, $\gamma=2a\mu_0$, 
which may be tuned to minimize: $[\sqrt{N}\sigma(\overline{\mathcal{O}}_k)]^2=
{{\rm var}({\mathcal{O}_k})\times2\tau_{{\rm int}}(\mathcal{O}_k)}$.
In this respect, it is important to note that in NSPT not only the integrated 
autocorrelation times $\tau_{{\rm int}}(\mathcal{O}_k)$ depend on the
parameters of the algorithm, but \emph{also} the variances 
${\rm var}({\mathcal{O}_k})$. This is so because, in general, the variances 
${\rm var}({\mathcal{O}_k})$ are not given by the perturbative expansion of any
correlation function of the theory. In particular, ${\rm var}({\mathcal{O}_k})$
does not correspond to the $2k$-order coefficient of $\langle\mathcal{O}^2\rangle-
\langle\mathcal{O}\rangle^2$. The independence of ${\rm var}({\mathcal{O}_k})$
on $\gamma$ is thus not guaranteed, and the exact dependence is a priori difficult
to infer. In the limit where $\gamma$ is kept fixed while $a\to0$, however, 
there is some theoretical control on this dependence. Specifically, assuming
$\langle\Obs\rangle|_{g_0^M}$ is properly renormalized, one can show that
all ${\rm var}({\mathcal{O}_k})$ are at most logarithmically divergent. Furthermore,
one can prove that $\tau_{{\rm int}}(\mathcal{O}_k)\propto a^{-2}$~\cite{Luscher:2011qa}.

%% file: sect3.tex
\section{The gradient flow coupling}
\label{sec:GFcoupling}

In order to study the viability of NSPT, we consider the 
computation of the GF coupling~\cite{Luscher:2010iy}. More precisely, 
we consider the definition advocated in~\cite{Fritzsch:2013je}, where
Schr\"odinger functional (SF) boundary conditions are imposed on the 
fields~\cite{Luscher:1992an}. Explicitly, these are given by,
\begin{equation}
  \label{eq:SFbc}
  U(x+\hat{k}L,\mu)=U(x,\mu),\quad
  U(x,k)|_{x_0=0,\,L}=\mathbbm{1},\quad
  k=1,2,3,
\end{equation}
where $L$ is the physical extent of the lattice in all four 
space-time directions, and $\hat{k}$ is the unit vector in 
the spatial direction $k$. A family of finite volume couplings can 
then be introduced as,
\begin{equation}
  \alpha_{\rm GF}(\mu)\propto
  \langle t^2 E_{\rm sp}(t,x)\rangle|_{x_0=L/2}
  \quad
  {\rm at}
  \quad
  \sqrt{8t}=cL\equiv1/\mu,
\end{equation}
where the constant $c$ defines the different renormalization schemes. 
The quantity $E_{\rm sp}(t,x)$, instead, corresponds to a given 
discretization of the spatial energy density of the flow field 
at flow time $t$ (cf. ref.~\cite{Fritzsch:2013je}). In the following 
we shall consider both the standard clover and plaquette definitions,
while the GF equations are discretized according to the Wilson flow
prescription~\cite{Luscher:2010iy}. 

Given these definitions, the goal is to determine the two-loop relation,
\begin{equation}
  \label{eq:GF2MSbar}
  \alpha_{\rm GF}(\mu) = \alpha_{\overline{\rm MS}}(\mu) + k_1\,\alpha^2_{\overline{\rm MS}}(\mu)
  +k_2\,\alpha^3_{\overline{\rm MS}}(\mu)+ {\rm O}(\alpha^4),
\end{equation}
in the continuum, where $\alpha_{\overline{\rm MS}}$ is the coupling 
in the $\overline{\rm MS}$ scheme. To this end, we first compute 
the relation between $\alpha_{\rm GF}$ and the bare coupling $\alpha_0=
g^2_0/(4\pi)$ with NSPT and then use the known two-loop relation between 
$\alpha_{\overline{\rm MS}}$ and $\alpha_0$ for our choice of lattice action~\cite{Luscher:1995np}. 
This gives us lattice approximants, $k_1(a/L)$, $k_2(a/L)$, of the continuum 
coefficients $k_1$, $k_2$, which need to be extrapolated to the continuum 
limit $a/L\to0$. The asymptotic form for $k_1(a/L)$, $k_2(a/L)$ close to 
the continuum limit is suggested by Symanzik's analysis~\cite{Symanzik:1979ph,
Symanzik:1982dy}, which in the present case gives,
\begin{equation}
  \label{eq:SymanzikAsympotic}
  k_l(a/L) \overset{a/L\to0}{\sim} k_l +\sum_{m=1}^{\infty}\sum_{n=0}^l c_{l,mn} 
  ({a/L})^m [\ln({L/a})]^n,\quad
  l=1,2.
\end{equation}
Observe that even though we are considering the pure SU(3) gauge theory, the 
presence of the SF boundary conditions (\ref{eq:SFbc}) introduces discretization 
effects proportional to odd powers of the lattice spacing. In particular,
O($a$) lattice artifacts, i.e $m=1$ terms in eq.~(\ref{eq:SymanzikAsympotic}),
are present. These can be removed by adding a O($a$) boundary counterterm to 
the action with an appropriately tuned coefficient $c_{\rm t}(g_0)$~\cite{Luscher:1992an}, 
which is actually known to the required two-loop accuracy. The
results that are presented in the following, however, have been obtained 
considering $c_{\rm t}$ only to tree-level i.e. $c_{\rm t}=1$. We then removed
the O($a$) contributions in $k_1(a/L)$ by an explicit analytic computation, 
while $k_2(a/L)$ is still affected by O($a$) lattice artifacts
corresponding to $m=1$, $n=0,1$, in eq.~(\ref{eq:SymanzikAsympotic}).

%% file: sect4.tex
\section{Results}
\label{sec:Results}

We generated ensembles with lattice sizes $L/a=10,12,16,20,24,32$.
The statistics we gathered was around 60'000 independent measurements 
for $L/a\leq 24$ and 80'000 for $L/a=32$. For the smaller lattices
these were equally divided in 6 different step-sizes in the range
$\delta\tau=0.1-0.238$. For $L/a=32$, instead, we only considered 
two step-sizes, $\delta\tau=0.126, 0.15$. The SMD equations were then 
integrated using a 4th-order symplectic scheme~\cite{Omelyan:2013}:
results are therefore correct up to O($\delta\tau^4$) errors. In fact, 
no statistically significant step-size errors were detected for 
any value of the step-size and lattice size considered. The chosen 
integrator hence performs really well. In addition, 
for lattices with $L/a\leq24$ we found perfect agreement with the 
results obtained using a Langevin implementation of NSPT based on 
the 2nd-order integrator of ref.~\cite{Bali:2013pla}. In conclusion, 
we are confident that within the statistical precision our results 
are not affected by step-size errors, and we can thus proceed discussing
their continuum extrapolation. Before doing so, we want to note that 
even though the Langevin results only showed very mild step-size errors, 
this set-up was not competitive for large lattices ($L/a>16$) due to 
rather long autocorrelations for $k_1(a/L),k_2(a/L)$. In the case of the
SMD algorithm, instead, the situation could be substantially improved by
a proper tuning of $\gamma$ (cf. Section {\ref{sec:NSPT}). Specifically,
we observed that the algorithm was most efficient when $\gamma$ was 
chosen in the range $\gamma=3-5$. Otherwise, the growth in the variances 
(autocorrelations) for smaller (larger) $\gamma$ values was significant,
especially for $k_2(a/L)$.

\begin{figure}[htbp]
  \centering
  \includegraphics[width=0.70\textwidth]{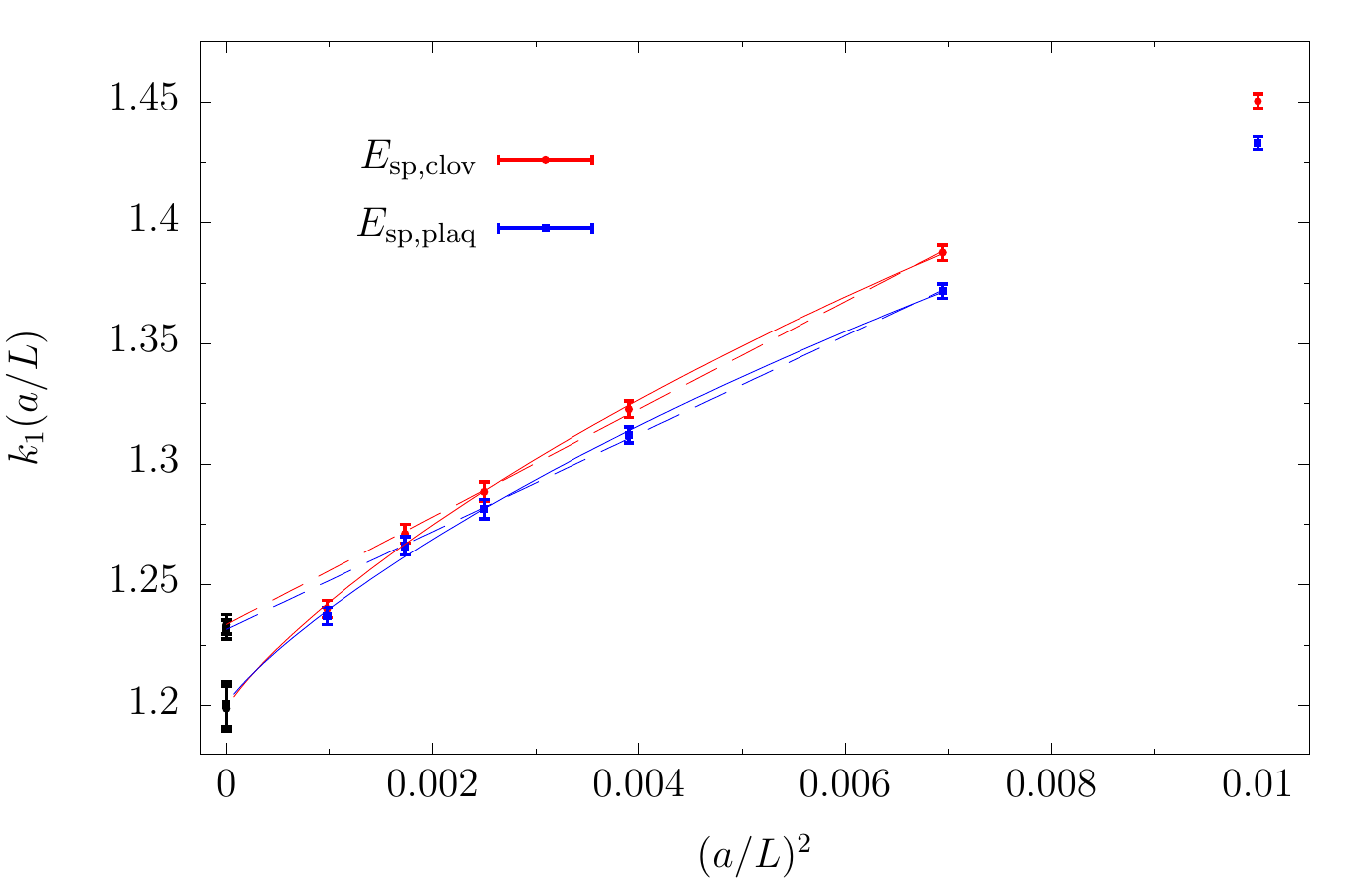}
  \includegraphics[width=0.70\textwidth]{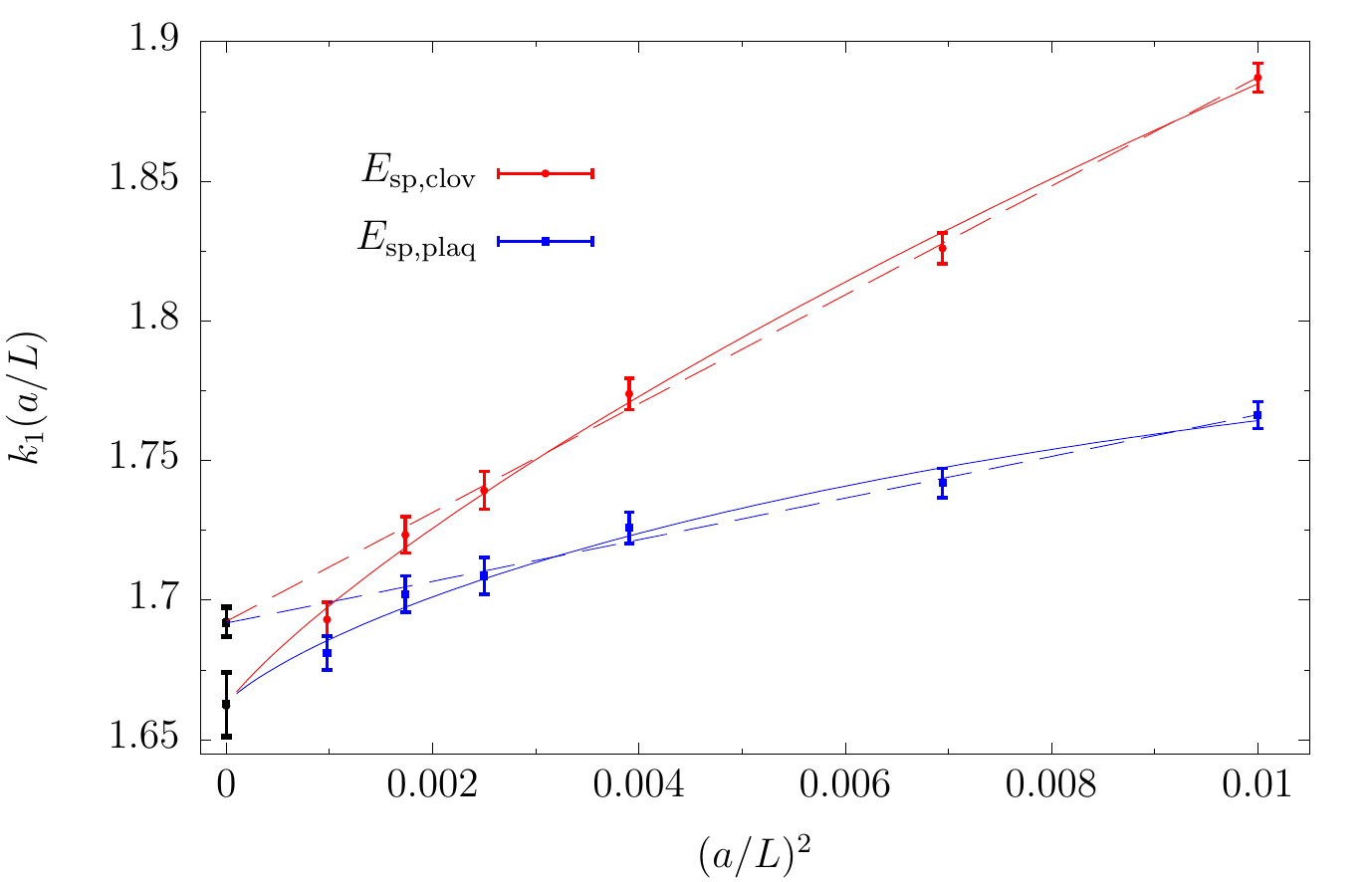}
  \caption{Continuum extrapolations for $k_1(a/L)$. Results for 
  $c=0.3$ (upper panel) and $c=0.4$ (lower panel) are shown, for 
  two different discretizations of the flow energy density.}
\label{fig:k1}
\end{figure}

In Fig.~\ref{fig:k1} we now present our results for $k_1(a/L)$
at $c=0.3$ and $0.4$. Two discretizations of the observable, namely clover 
and plaquette, are plotted. Focusing on the results for $c=0.3$, we first 
note that the statistical errors on $k_1(a/L)$ are around $4\cdot10^{-3}$. 
In the plot we then show two different extrapolations to the continuum limit. 
In the first extrapolation (solid line), lattices with $L/a\geq12$ are 
fitted to the asymptotic form (\ref{eq:SymanzikAsympotic}) including the
leading terms $m=2$, $n=0,1$. This results in a $\chi^2/{\rm d.o.f.}\sim1$.
A first important  observation is that O($a^2$) effects are sizable, i.e. 
on the level of 0.1 at $L/a=10$. Secondly, the $m=2$, $n=1$, term is crucial
to obtain a good fit of the data. In the second extrapolation (dashed line),
instead, we considered lattices with $L/a\geq 12$ but excluded $L/a=32$. 
In this case the data can be very well described by a pure $(a/L)^2$ term 
over the whole range of lattice sizes. The continuum extrapolated value so
obtained has smaller statistical errors, but is many standard deviations away
from the result of the previous fit. 

The results for $c=0.4$ exhibit qualitatively the same features,
although the statistical errors of $k_1(a/L)$ are now $\sim7\cdot10^{-3}$
and the two discretizations of the observable show rather different 
lattice artifacts. On the other hand, cutoff effects are generally smaller 
than for $c=0.3$, and we thus included the $L/a=10$ data into the fits as well. 

Given the above observations, we conclude that the extraction of continuum
perturbative coefficients from NSPT data is rather delicate due to
the presence of the logarithmic corrections to the continuum scaling. These
are significant in the present case at a level of precision just below $10^{-2}$.
Larger lattices are certainly needed to better constrain the observed behavior,
and extract definite and reliable continuum results.

Accurate results are more difficult to obtain for the two-loop coefficient $k_2$
than for $k_1$. The statistical errors turn out to be about 10 times larger in 
this case and the extrapolation to the continuum limit is complicated by the
presence of further terms in the fit function (cf.~eq.~(\ref{eq:SymanzikAsympotic})). 
A reliable analysis of the data for $k_2$ therefore has to wait for the ongoing
simulations of larger (as well as some smaller) lattices to be completed.

%% file: sect5.tex
\section{Conclusions}
\label{sec:Conclusions}

In this contribution we investigated the possibility of using NSPT 
to compute the perturbative expansion of physical quantities in the 
\emph{continuum} theory. To this end, we considered a  non-trivial
case: the determination of the two-loop matching between the GF 
coupling in finite-volume with SF boundary conditions and the 
$\overline{{\rm MS}}$ coupling, in the pure SU(3) gauge theory. 
The use of the SMD in place of the Langevin algorithm proves to be
beneficial in this context and allows statistically precise results
to be obtained near the continuum limit with a significantly reduced
computational effort. In this respect, we note that an \emph{absolute}
uncertainty on the matching coefficients of $\sigma(k_1)\sim 10^{-2}$ 
and $\sigma(k_2)\sim 10^{-1}$, would imply a \emph{relative} error on 
the determination of $\alpha_{\overline{\rm MS}}(m_Z)$ around $0.2\%$
i.e. well below the error of state of the art non-perturbative 
determinations~\cite{Brida:2016flw,DallaBrida:2016}. 

Taking the continuum limit of the calculated coefficients can be challenging
in view of the statistical uncertainties and the fact that the dependence
on the lattice spacing is not simply power-like. Reliable extrapolations
probably require O($a$)-improvement up to the order in the coupling 
considered and certainly accurate data over a significant range of lattice 
sizes extending up to some fairly large ones.

The inclusion of the quark fields in the SMD algorithm is in principle 
straightforward and is not expected to slow down the simulations by a large
factor~\cite{DiRenzo:2004ge}. Different implementations are however possible,
whose viability and efficiency will need to be determined.

%% file: acknw.tex
\section{Acknowledgments}

M.D.B. has benefited from the pleasant collaboration with 
Marco Garofalo, Dirk Hesse, and Tony Kennedy on related investigations.
He is also grateful to Alberto Ramos, Stefan Sint, and Rainer Sommer, 
for their valuable comments. The authors thank LRZ for the allocated 
computer time under the project id: \texttt{pr92ci}, as well as the 
computer centers at CERN and DESY -- Zeuthen, for their precious 
support and resources.

%% file: bibl.tex
\bibliographystyle{JHEP}
\bibliography{bibl}

%% file: main.bbl
\providecommand{\href}[2]{#2}\begingroup\raggedright\begin{thebibliography}{10}

\bibitem{Luscher:2010iy}
M.~L{\"u}scher, {\em JHEP} {\bf 1008} (2010) 071. [Erratum: {\em JHEP} {\bf 03} (2014) 092.]

\bibitem{Luscher:2011bx}
M.~L{\"u}scher and P.~Weisz, {\em JHEP} {\bf 1102} (2011) 051.

\bibitem{Luscher:2013vga}
M.~L{\"u}scher, {\em PoS} {\bf LATTICE2013} (2014) 016.

\bibitem{Fodor:2012td}
Z.~Fodor et. al., {\em JHEP} {\bf 11} (2012) 007.

\bibitem{Fritzsch:2013je}
P.~Fritzsch and A.~Ramos, {\em JHEP} {\bf 1310} (2013) 008.

\bibitem{Suzuki:2013gza}
H.~Suzuki, {\em PTEP} {\bf 2013} (2013) 083B03. [Erratum: {\em PTEP} {\bf 2015} (2015) 079201.]

\bibitem{Bribian:2016}
E.~I. Bribian and M.~Garc\'ia~P\'erez, {\em PoS} {\bf LATTICE2016} (2017) 371.

\bibitem{Harlander:2016vzb}
R.~V. Harlander and T.~Neumann, {\em JHEP} {\bf 06} (2016) 161.

\bibitem{DiRenzo:1994sy}
F.~Di~Renzo, E.~Onofri, G.~Marchesini, and P.~Marenzoni, {\em Nucl. Phys.} 
   {\bf B426} (1994) 675--685.

\bibitem{DiRenzo:2004ge}
F.~Di~Renzo and L.~Scorzato, {\em JHEP} {\bf 10} (2004) 073.

\bibitem{Brida:2013mva}
M.~Dalla~Brida and D.~Hesse, {\em PoS} {\bf LATTICE2013} (2013) 326.

\bibitem{Brida:2015d}
M.~Dalla~Brida, M.~Garofalo, and A.~D. Kennedy, {\em PoS} {\bf LATTICE2015} (2015) 309.

\bibitem{Horowitz:1985kd}
A.~M. Horowitz, {\em Phys. Lett.} {\bf B156} (1985) 89.

\bibitem{Horowitz:1986dt}
A.~M. Horowitz, {\em Nucl. Phys.} {\bf B280} (1987) 510.

\bibitem{Luscher:2011kk}
M.~L{\"u}scher and S.~Schaefer, {\em JHEP} {\bf 07} (2011) 036.

\bibitem{Luscher:2011qa}
M.~L{\"u}scher and S.~Schaefer, {\em JHEP} {\bf 04} (2011) 104.

\bibitem{Hesse:2013lat}
M.~Brambilla et al., {\em PoS} {\bf LATTICE2013} (2013) 325.

\bibitem{Luscher:1992an}
M.~L{\"u}scher, R.~Narayanan, P.~Weisz, and U.~Wolff, {\em Nucl. Phys.} {\bf B384} (1992) 168--228.

\bibitem{Luscher:1995np}
M.~L{\"u}scher and P.~Weisz, {\em Nucl. Phys.} {\bf B452} (1995) 234--260.

\bibitem{Symanzik:1979ph}
K.~Symanzik, {\em NATO Sci. Ser.} {\bf B59} (1980) 313--330.

\bibitem{Symanzik:1982dy}
K.~Symanzik, {\em J. Phys. Colloq.} {\bf 43} (1982) C3 254.

\bibitem{Omelyan:2013}
I.~P. Omelyan, I.~M. Mryglod, and R.~Folk, {\em Comp. Phys. Commun.} {\bf 151} (2003) 272.

\bibitem{Bali:2013pla}
G.~S. Bali, C.~Bauer, A.~Pineda, and C.~Torrero, {\em Phys. Rev.} {\bf D87} (2013) 094517.

\bibitem{Brida:2016flw}
M.~Dalla~Brida et. al., {\em Phys. Rev. Lett.} {\bf 117} (2016), no.~18 182001. 

\bibitem{DallaBrida:2016}
M.~Dalla~Brida et. al., {\em PoS} {\bf LATTICE2016} (2017) 197.

\end{thebibliography}\endgroup
